\documentclass[twocolumn,english,aps,pra,superscriptaddress,letterpaper]{revtex4-1}

\usepackage{amsmath,amssymb,graphicx}
\usepackage[colorlinks=true,bookmarks=false,citecolor=blue,urlcolor=blue]{hyperref} 
\usepackage{mathtools}
\usepackage{lineno}
\newcommand{\xzpf}{x_\text{zpf}}
\newcommand{\meff}{m_\text{eff}}

\newcommand{\wring}{w_\text{ring}}
\newcommand{\omegam}{\Omega_\text{m}}
\DeclarePairedDelimiter\bra{\langle}{\rvert}
\DeclarePairedDelimiter\ket{\lvert}{\rangle}
\newcommand{\vin}{v_\text{in}}
\newcommand{\vout}{v_\text{out}}
\newcommand{\tm}{\mathbf{M}}
\newcommand{\mean}[1]{\langle #1 \rangle}
\newcommand{\ten}[1]{\mathbf{#1}}

\newcommand{\kB}{k_\text{B}}
\newcommand{\omegamod}{\Omega_\text{mod}}
\newcommand{\kappae}{\kappa_\text{e}}

\newcommand{\ain}{a_\text{in}}

\newcommand{\omegap}{\omega_\text{p}}

\begin{document}

\title{Hybrid confinement of optical and mechanical modes in a bullseye optomechanical resonator}

\author{Felipe G. S. Santos}
\author{Yovanny A. V. Espinel}
\author{Gustavo O. Luiz}
\author{Rodrigo S. Benevides}
\author{Gustavo S. Wiederhecker}\email{gustavo@ifi.unicamp.br}
\author{Thiago P. Mayer Alegre}\email{alegre@ifi.unicamp.br} \email{\\http://nanophoton.ifi.unicamp.br}
\affiliation{Applied Physics Department, “Gleb Wataghin” Physics Institute, University of Campinas - UNICAMP,  13083-859 Campinas, SP, Brazil}

\begin{abstract}
Optomechanical cavities have proven to be an exceptional tool to explore fundamental and technological aspects of the interaction between mechanical and optical waves. Such interactions strongly benefit from cavities with large optomechanical coupling, high mechanical and optical quality factors, and mechanical frequencies larger than the optical mode linewidth, the so called resolved sideband limit. Here we demonstrate a novel optomechanical cavity based on a disk with a radial mechanical bandgap. This design confines light and mechanical waves through distinct physical mechanisms which allows for independent control of the mechanical and optical properties. Our device design is not limited by unique material properties and could be easily adapted to allow large optomechanical coupling and high mechanical quality factors with other promising materials. Finally, our demonstration is based on devices fabricated on a commercial silicon photonics facility, demonstrating that our approach can be easily scalable.
\end{abstract}

\maketitle 
\newcommand{\nocontentsline}[3]{}
\newcommand{\tocless}[2]{\bgroup\let\addcontentsline=\nocontentsline#1{#2}\egroup}


Optomechanical microcavities simultaneously confine optical and mechanical modes. The interaction between these confined modes has proven to be a very rich field of study for basic science such as macroscopic quantum phenomena~\cite{Chan:2011dy, Chen:2013dh}, quantum simulation of condensed matter phenomena~\cite{Schmidt:2015bt, Schmidt:2015jk}, topological phase pattern-formation of coupled oscillators~\cite{Peano:2015bn} and non-classical states of light~\cite{SafaviNaeini:2014cx, Riedinger:2016cl}. Technological questions have been addressed by optomechanical devices as well, with applications including weak-force sensing~\cite{Forstner:2014gx, Forstner:2012ci, Sun:2012je}, ultra-sensitive accelerometers~\cite{Krause:2012cfa}, radio-frequency sources~\cite{Luan:2014js}, multi-mode~\cite{Grutter:2015km} and synchronous~\cite{Zhang:2012ks} oscilators, reconfigurable optical filters~\cite{Deotare:2012cp} and hybrid systems~\cite{Balram:2016eu}.

Regardless of the context, the optomechanical interaction often benefits from a large optomechanical coupling rate, low optical and mechanical dissipation rates (high-$Q$s) and the so-called resolved sideband limit, which occurs when the mechanical mode frequency is larger than the optical mode linewidth. The optomechanical coupling rate is denoted by $g_0$ and measures the optical cavity frequency shift induced by a mechanical mode with an amplitude equivalent to the quantum harmonic oscillator at ground-state.

The challenge in simultaneously achieving an optimum combination of these properties is that the optical and mechanical optimization are often constrained to each other. One promising platform is based on the optomechanical crystal cavity, where both waves are confined by bandgaps in two-dimensional periodic structures. So far, however, only a single microcavity design has been demonstrated that can simultaneously confine optical and mechanical modes~\cite{SafaviNaeini:2014gh}. These devices were fabricated using the traditional approach of a direct write electron beam lithography. However, for massive fundamental studies and applications is desirable to migrate the fabrication process of these devices to commercial CMOS-compatible facilities, where the photo-lithography process considerably reduces time and cost when compared to the traditional approach.
 
\begin{figure*}[t]
	\centerline{\includegraphics[scale=0.9]{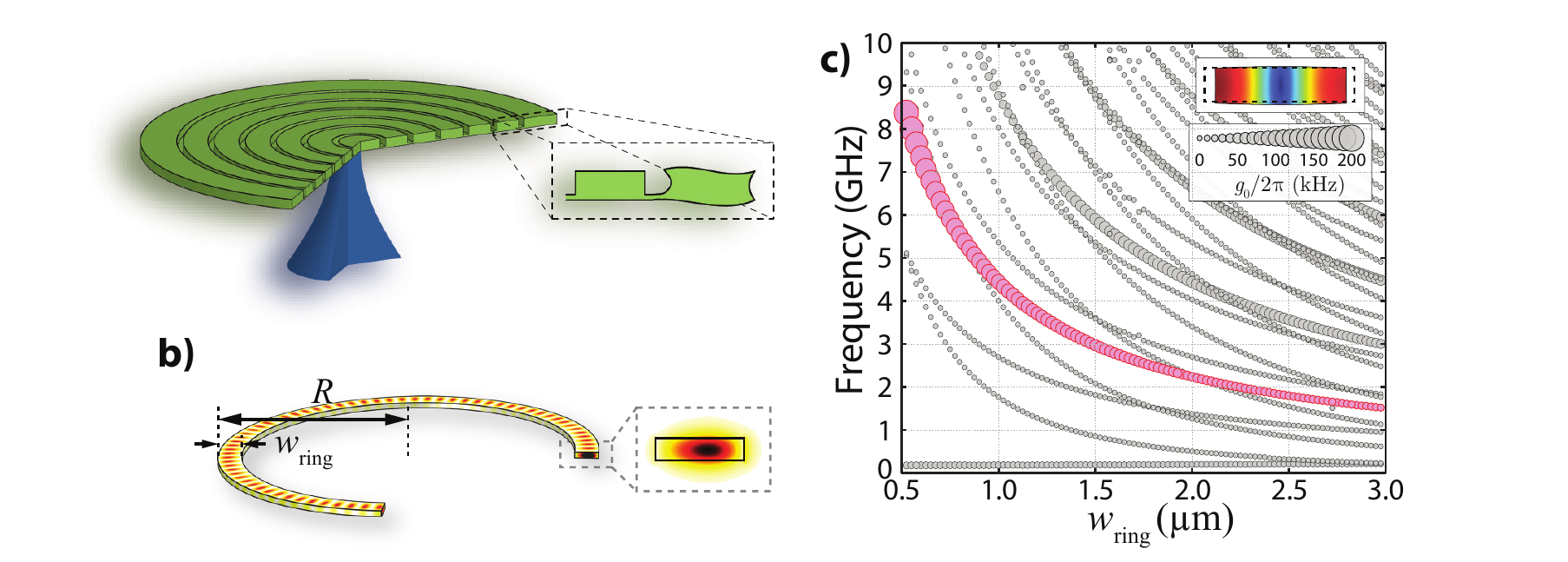}}
	\caption{\textbf{Optomechanical ring cavity.} 
	\textbf{a)} Schematic of a bullseye optomechanical cavity (green) supported by a central pedestal (blue). The inset illustrates the mechanical motion localized at the disk's edge. 
	\textbf{b)} Finite-element simulation (FEM) of the electric field intensity for the fundamental transverse-electric (TE) mode of a floating ring resonator, which is defined by the external radius $R$ and the ring width $\wring$. \emph{Inset}: darker (lighter) colors represents larger (lower) electric field intensity.
	\textbf{c)} Mechanical dispersion of a ring as a function of $\wring$ for $R=8~\mu\text{m}$. The marker size represents the optomechanical coupling, $g_0$, between each mechanical mode to the TE fundamental optical mode shown in \textbf{b)}. Red dots represent the first order mechanical breathing mode. \emph{Inset}: cross-section of the first order breathing mode for $\wring=1~\mu\text{m}$. Red (blue) color accounts for larger (smaller) total displacement field.}
	\label{fig:ring}	
\end{figure*}
Here we propose a new optomechanical device based on a silicon microdisk with a circular mechanical grating fabricated on a CMOS-compatible foundry (Fig.~\ref{fig:ring}a). The optical waves are confined at the disk edge in whispering gallery-like modes due to total internal reflection, whereas the mechanical modes are confined as an edge state lying within a phononic bandgap of the circular grating. We show that these two unrelated confining mechanisms relax the need for simultaneous optical and mechanical bandgaps, while keeping both optical and mechanical losses minimal. These relaxed requirements are encouraging for addressing scalable applications of optomechanical cavities.

Using both numerical simulations and measurements we demonstrate a tailorable mechanical bandgap up to several GHz within an experimentally accessible range of geometrical parameters. The calculated optomechanical coupling rate is rather high, $g_0/2\pi\approx100$ kHz, which is about one order of magnitude larger than regular silicon microdisks with the same radius~\cite{Aspelmeyer:2014ce}. Mechanical radiation losses are also inhibited by the mechanical bandgap, which allows for a high mechanical $Q$-factor, essentially limited by material losses.

In order to gain some insight on our final device design, we start by analyzing a floating ring resonator (Fig.~\ref{fig:ring}b). This idealized optomechanical structure benefits from the highly co-localized optical and mechanical modes defined by its geometrical constraints~\cite{VanLaer:2015jk}. In this case, mechanical losses are reduced to material losses while optical losses are due to material and to radiation (bending) loss.
Two main features are captured by this simple model, the ability to tailor the mechanical frequency from a couple to tens of GHz, while raising the optomechanical coupling rate from tens to hundreds of kHz. One can also easily evaluate the optomechanical coupling rate, $g_0$, following a perturbation theory approach for such structure:
\begin{equation}
g_0=-\frac{\omega}{2}\frac{\bra{\vec{E}}\Delta\epsilon\ket{\vec{E}}}{\bra{\vec{E}}\epsilon\ket{\vec{E}}},
\label{eq:g0}
\end{equation}
where the inner products ratio between the unperturbed electric field distribution ($\vec{E}$) measures how much the optical mode resonance frequency, $\omega$, varies when the dielectric constant changes from $\epsilon$ to $\epsilon+\Delta\epsilon$ because of the mechanical mode. Two main contributions are taken into account for evaluating $\Delta\epsilon$: one caused by the change in the dielectric constant spatial distribution due to the deformation of the boundaries~\cite{Johnson:2002jd} and another coming from strain-induced modifications of the refractive index (photoelastic effect)~\cite{Chan:2012iy}. In Fig.~\ref{fig:ring}c we show the mechanical frequency and optomechanical coupling dependence of the first few mechanical modes on the ring width. 
We also highlight the lowest order breathing mode (red dots) in Fig.~\ref{fig:ring}c; larger markers account for larger $g_0=g_0^\text{MB}+g_0^\text{PE}$, where $g_0^\text{MB}$ and $g_0^\text{PE}$ account for the moving boundary and photo-elastic contributions respectively. Such high $g_0$ is dominated mainly by the photoelastic contribution (see Supplemental material). As an example, a device with a ring radius of $R=8~\mu$m and ring width of $w_\text{ring}=1~\mu$m would have a net optomechanical coupling rate of $g_0/2\pi=88$~kHz at a mechanical frequency of $4.5$~GHz (throughout the text silicon-on-insulator device layer was kept fixed and equal to $220$~nm). Both values are one order of magnitude larger than the radial-breathing modes in a Si disk with the same radius~\cite{Aspelmeyer:2014ce} (not shown). Despite idealized, this system gives an upper boundary for the optomechanical coupling rate as well as an insight for the edge state mechanical mode on our structure.

\begin{figure*}[t!]
	\centerline{\includegraphics[scale=0.9]{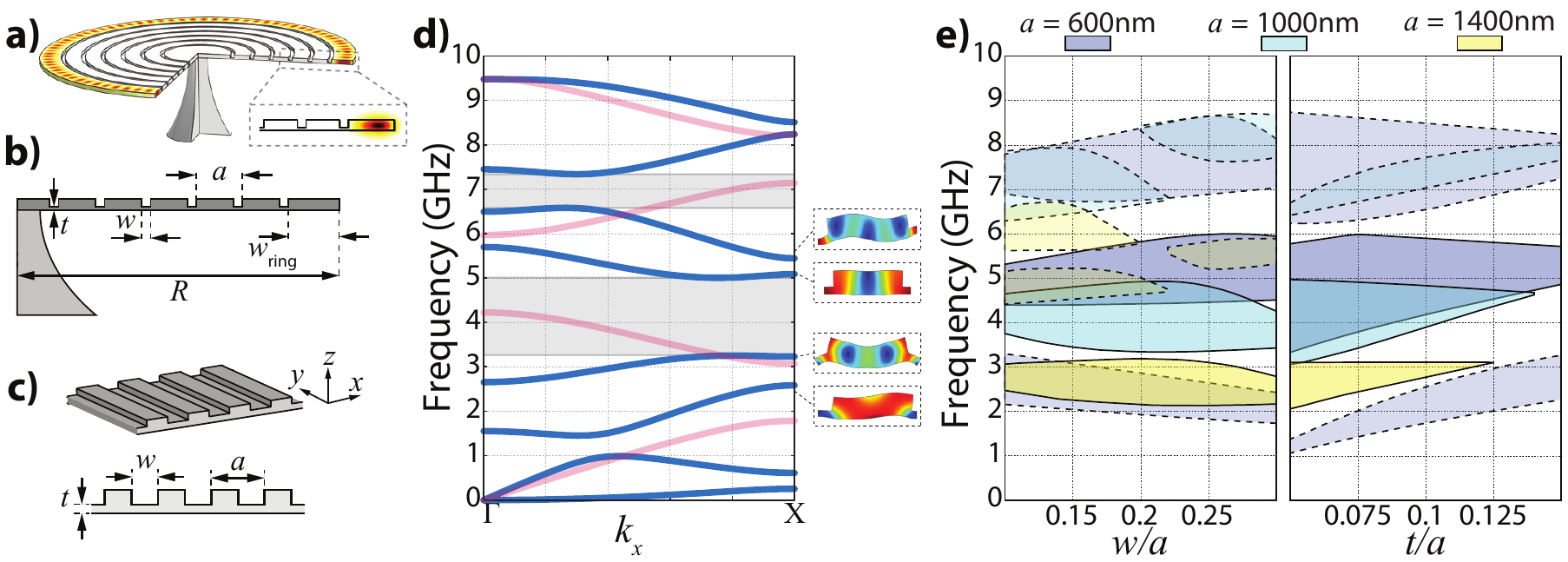}}
	\caption{\textbf{1D phononic crystal} 
	\textbf{a)} Bullseye axisymmetric FEM simulation of the electric field intensity for the first order TE-whispering gallery (inset). Simulation parameters are $a=1~\mu$m, $w=200$~nm, $t=70$~nm, $\wring=1.1~\mu$m and, $R=8~\mu$m.
	\textbf{b)} Each geometrical parameters of the bullseye disk controls distinct physical properties: $R$ defines the optical frequency and free-spectral range, $\wring$ defines the mechanical frequency while the grating parameters $a$, $w$ and $t$ determine the phononic bandgap. 
	\textbf{c)} 1D periodic phononic crystal approximation for the  circular grating show in \textbf{b)}. 
	\textbf{d)} The blue (red) lines are the mechanical bands of the 1D phononic crystal for $x$-polarized or $z$-polarized ($y$-polarized) modes for $a=1~\mu$m, $w=200$~nm and $t=70$~nm. The insets show the mechanical deformation for the floquet-modes at the band-edge ($X$-point) of selected edge states. Red (blue) colors accounts for larger (smaller) displacement field. 
	\textbf{e)} Bandgap maps for the 1D phononic crystals for $x$ an $z$-polarized waves. The colored areas correspond to regions within bandgaps for different lattice period $a$. The regions delimited by solid lines correspond to bandgaps between the two modes around 3~GHz and 5~GHz shown in \textbf{d)}. Left (Right) bandgap maps as a function of $w/a$  ($t/a$) for fixed $t=70$~nm ($w/a=0.25$). All dimensions are compatible with Si-photonics foundry based processes.}
	\label{fig:crystal}
\end{figure*}

\begin{figure*}[hbt!]
	\centerline{\includegraphics[scale=0.9]{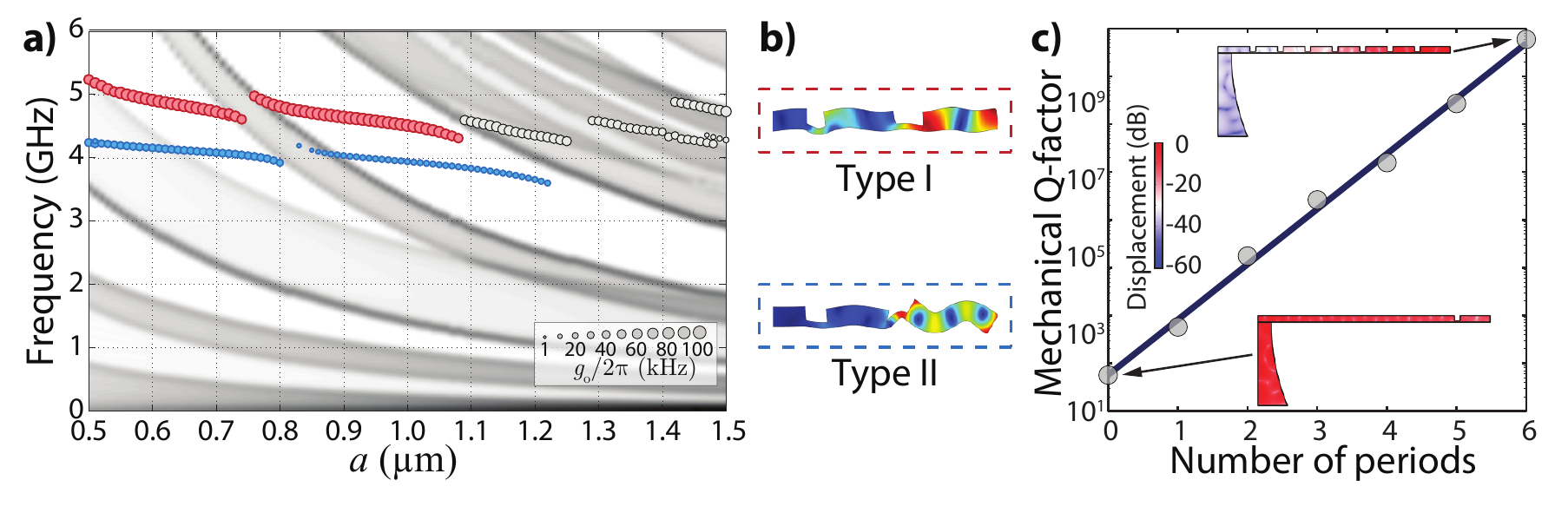}}
	\caption{\textbf{Optomechanical bullseye disk.} 
	\textbf{a)} FEM simulations for the mechanical dispersion of a bullseye disk (solid dots) as a function of the periodicity $a$ for $\wring=1.1~\mu$m, $w/a=0.25$ and $t=70$~nm. The gray tone shades are proportional to the mechanical density of states (DOS) of the corresponding 1D phononic crystal, where darker regions are related to higher DOS within the mechanical grating. Only the two largest $g_0$ modes are shown. Symbol sizes are proportional to the total optomechanical coupling rate. 
	\textbf{b)} Type-I and type-II displacement profile of the confined mechanical modes for $a=1~\mu$m. type-I is the first order breathing mode at $4.2$~GHz; type-II is a high order flapping mode at $3.8$~GHz. 
	\textbf{c)} Simulated mechanical radiation loss quality factor for the type-I mode as a function of the number of grating periods. The solid line is a fit for the simulated points with an exponential decay curve. The insets show the normalized mechanical displacement in log scale for zero (no anchor clamping suppression) and six grating periods ($\approx60$~dB of isolation from the clamping region).}
	\label{fig:bullseye}
\end{figure*}
In order to confine optical and mechanical modes together in space, we mimic a floating ring-like resonator by using a \emph{phononic shield} in the form of a bullseye radial Bragg gratting (see Fig.~\ref{fig:crystal}~a-b)). Such structure have successfully being used to confine photonic~\cite{Labilloy:1998gx, Yariv:2003et, Schonenberger:2009kz} and plasmonic~\cite{Lezec:2002bq, Jun:2011el, Yi:2014fn} modes. The mechanical waves reflected by this circular grating can be understood using the one-dimensional approximation shown in Fig.~\ref{fig:crystal}c. This approximation is quite precise since the interfaces interacting with a cylindrical wave traveling through the circular grating are similar to those interacting with a plane wave in a linear crystal (see Supplemental Material). The one-dimensional approximation then ease the understand and design of the bullseye Bragg grating.

We are interested in creating a phononic bandgap that confines the radial breathing mechanical modes at the disk edge. These modes arise from longitudinal waves (L) propagating along the radial direction, which are mixed with shear-vertical (SV) waves due to reflection at the free-boundaries ~\cite{Auld:1990ub}. Therefore, a phononic shield with a partial bandgap for L and SV waves is enough to confine to the ring-like breathing mode. Fig.~\ref{fig:crystal}d shows a typical band structure of the 1D phononic crystal; blue lines represent $x$-polarized (L) or $z$-polarized (SV) modes, whereas the red lines represent $y$-polarized (Shear-Horizontal, SH) modes. A partial bandgap (gray shades) is then formed around between $3.2$~GHz and $5$~GHz.

To confine the first order mechanical breathing mode at any desired frequency (chosen via $\wring$), we further investigate the dependence of the partial bandgaps on the grating parameters, $a$, $w$ and $t$ (Fig.~\ref{fig:crystal}e). We also restrict the design parameters within the range accessible through Si-photonics foundries. The left panel on Fig.~\ref{fig:crystal}e shows the main bandgaps (colored regions) as a function of $w/a$ for different values of $a$ keeping $t=70$~nm, while the right graph shows the dependence of such bandgap on $t/a$ while maintaining $w/a=0.25$. 
As expected, raising $a$ both shrinks the total bandgap and moves it towards lower frequencies. It is also worth noting that the bandgaps around the region of interest widens as $t/a$ is decreased, a consequence of the large acoustic velocity contrast between the thicker and thinner regions.

One can easily find a suitable design of a bullseye disk with larger optomechanical coupling for a given mechanical frequency using the bandgap maps on Fig.~\ref{fig:crystal}e, and the relation between the mechanical frequency and the ring width from Fig.~\ref{fig:ring}c. For example, choosing a $4.0$~GHz mechanical mode frequency, we determine from Fig.~\ref{fig:ring}c that the ring width should be $\wring=1.1~\mu$m and from the bandgap maps that $a=1~\mu$m with $w/a=0.2$ and $t/a=0.07$.

\begin{figure*}[t]
	\centerline{\includegraphics[scale=0.9]{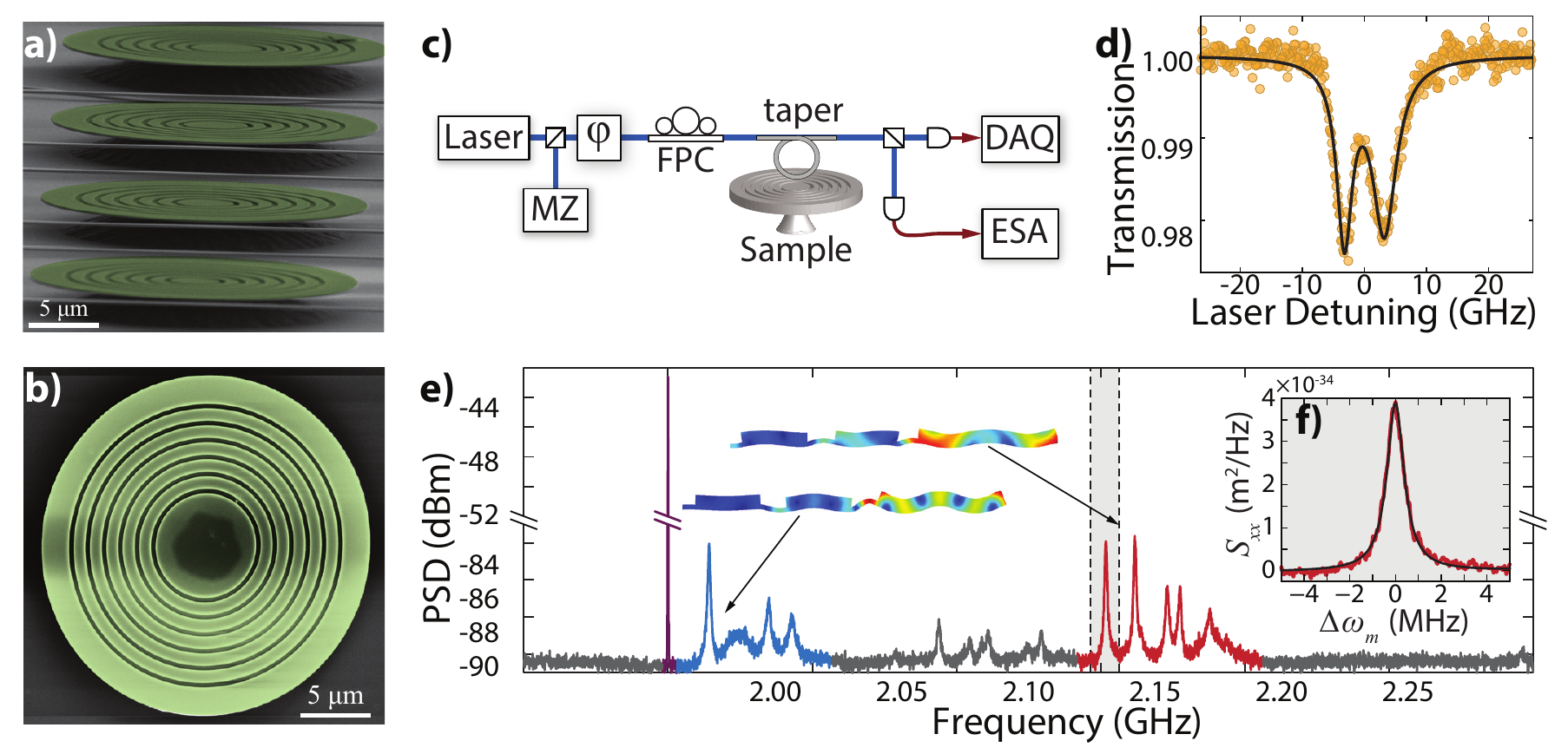}}
	\caption{\textbf{Mechanical modes.}
		\textbf{a)} and \textbf{b)}  Scanning electron microscope (false color) images of typical bullseye cavities from side and top respectively.
		\textbf{c)} Experimental setup. DAQ: Data Acquisition System (records optical spectra); ESA: Electrical Spectrum Analyzer (records mechanical spectra); $\varphi$: electro-optical phase modulator; MZ: Mach-Zehnder interferometer (wavelength calibration); FPC: Fiber Polarization Controller.
		\textbf{d)} Typical optical resonance used to address the bullseye's mechanical modes ($Q_\text{opt}\approx4\times10^4$) and exhibiting forward-backward splitting due to surface roughness back-scattering.
		\textbf{e)} The noise spectrum is composed by the external phase modulator tone (pink) and several peaks from to the thermal motion. The higher frequency family of peaks (red) is attributed to type-I modes whereas the lower frequency one (blue) has to do with type-II mechanical modes.
		\textbf{f)} Fitting the noise spectrum to the typical thermal mechanical response allows us to determine the mechanical mode frequency, quality factor and, along with knowledge of the external phase modulation, $g_0$. For $a=1400$~nm, the highest peak belonging to the type-I family shows $g_0/2\pi=23.1\pm0.2$~kHz and $Q_\text{m}=2360\pm40$.
		}
	\label{fig:experiments}
\end{figure*}

\begin{figure*}[t]
	\centerline{\includegraphics[scale=0.9]{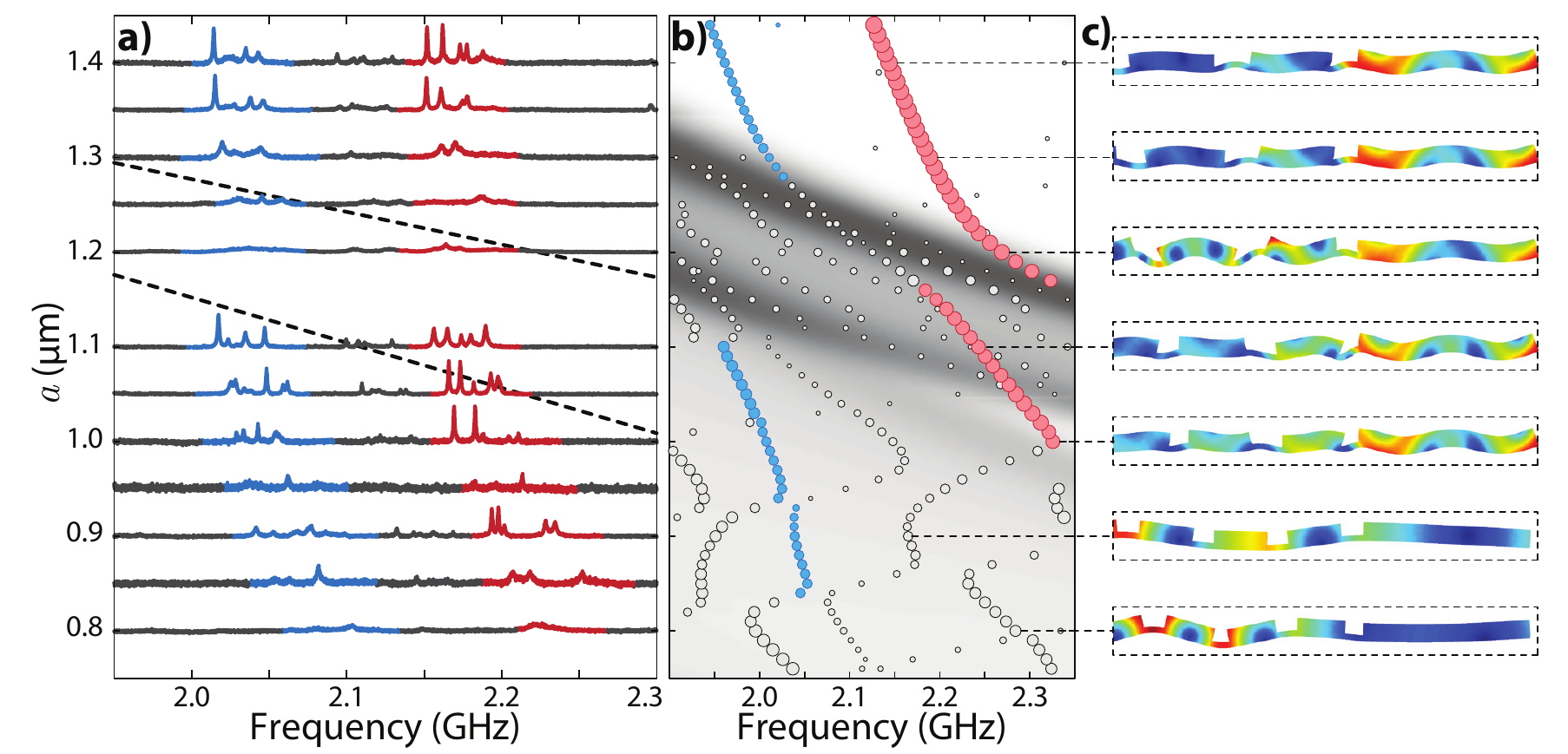}}
	\caption{\textbf{Bullseye experimental demonstration.}
		\textbf{a)} Mechanical spectra as a function of $a$ for fixed $\wring=2~\mu$m, $w/a=0.2815$, $t=70$~nm and silicon layer of $220$~nm. The vertical axis is normalized in order to account only for the optomechanical response (see Methods). The dashed lines accounts for first and second band edge modes displacement profiles shown in Fig.~\ref{fig:crystal}~\textbf{d)}.
		\textbf{b)} Simulated mechanical modes for an axisymmetric structure with the same parameters as those in \textbf{a)}. Larger (smaller) marker size accounts for larger (smaller) optomechanical coupling rate to the first order TE mode shown on Fig.~\ref{fig:crystal}~\textbf{a)}. The red (blue) color is a guide to the mechanical modes of type-I (type-II). The gray tone shades are proportional to the mechanical DOS for the corresponding 1D phononic crystal, where darker (lighter) regions are related to higher (smaller) density of states (DOS) within the mechanical grating.
		\textbf{c)} Calculated mechanical mode displacement profile for selected type-I modes. Red (blue) colors represent larger (zero) mechanical displacement.}
	\label{fig:grating}
\end{figure*}

In order to further evaluate this optimized design, we use (axisymmetric) Finite Element Method (FEM) simulations for the optical and mechanical modes of the whole bullseye structure to calculate the optomechanical coupling rate, $g_0$, and the radiation limited mechanical quality factor ($Q_m$). The results are shown in Fig.~\ref{fig:bullseye}a superimposed on a mechanical density of states map of the corresponding linear crystal; higher density of states (DOS) are represented by darker regions. For such structure two mechanical mode types give notoriously high optomechanical coupling of $g_0^\text{(I)}/2\pi=77$~kHz and $g_0^\text{(II)}/2\pi=30~$kHz. Similarly to the floating ring, these couplings are dominated by the photoelastic effect due to the large strain within the outermost ring overlap with the localized fundamental transverse electric (TE) optical mode.

The type-I mode shown on Fig.~\ref{fig:bullseye}b is related to the first order breathing mode of the ring structure when it's frequency is within the bandgap of the analogue linear crystal. On the other hand, the type-II mode is related to a high order flapping mode. In the floating ring structure such mode has nearly zero optomechanical coupling due to the opposite parity between the optical and mechanical modes. However, in the bullseye disk this parity is broken due to the position of the slab connecting each grating ring, resulting in a finite $g_0$ for these flapping modes.

Since radial waves within the bandgap cannot propagate towards the pedestal, anchor losses are drastically reduced. As expected for a phononic bandgap, the mechanical quality factor grows exponentially with the number of periods in the grating, as shown in FEM simulation (using  perfect matched layers) results in Fig.~\ref{fig:bullseye}c.

\tocless{\section*{Demonstration}}
We experimentally demonstrate the bullseye design in devices fabricated by a commercial foundry (Fig.~\ref{fig:experiments}a). Using a CMOS commercial foundry has several advantages towards coupled optomechanical cavity arrays as well as implementing high volume and low cost on-chip circuitry and sensing. The fabrication is based on a deep-UV optical lithography (see \emph{Methods}) with nominal resolution of $130$~nm. 

In order to evaluate the effects of the bullseye grating structure we fabricated devices with $\wring=1.5~\mu$m and $2.0~\mu$m, corresponding to first order mechanical breathing modes around $3.0$~GHz (see Fig.~\ref{fig:fano2}) and $2.0$~GHz respectively, while varying the bullseye pitch from $a=650$~nm to $1450$~nm. We also change the grating filling factors from $w/a=0.25$ up to 0.31.

The devices are probed by a tunable laser that is evanescently coupled to the cavity through a tapered optical fiber (Fig.~\ref{fig:experiments}c). A lorentzian fit to the DC optical transmission signal yields intrinsic optical quality factors of $4\times10^4$ ($\kappa_i/2\pi\approx5$~GHz), as shown in Fig.~\ref{fig:experiments}d. By changing the distance between the fiber taper to the optical cavity we could optimizing the optical coupling and achieve almost critical coupling (not shown, $\kappa_e\approx\kappa_i$). We also used a similar technique to ensure the excitation of edge-localized optical modes,as described in \emph{Methods}. 

Room temperature thermal fluctuations excite the mechanical modes that modulate the phase of the intra-cavity optical field. The cavity resonance dispersion converts the phase-modulation into an amplitude modulation of the transmitted light, which can be measured using a radio-frequency (RF) spectrum analyzer. Despite the high mechanical frequencies of the fabricated device, the rather low optical quality factor (limited by the foundry lithography) place the optomechanical cavity barely in the resolved sideband limit ($\kappa/2\approx\omega_m$), therefore, the laser is tuned to the maximum slope of the optical mode transmission to ensure the most efficient conversion from phase to amplitude modulation. In order to calibrate the optomechanical coupling rate, $g_0$, the input laser beam is modulated at a frequency  close the mechanical resonance with a calibrated phase modulator.~\cite{Gorodetsky:2010jd}. Fig.~\ref{fig:experiments}e shows a typical RF spectrum around the mechanical modes of a device with $\wring=2.0˜\mu$m and $a=1400$~nm. The modes around $2.15$~GHz (red) are identified as the type-I mechanical modes whereas those around $2.00$~GHz (blue) are type-II modes. The RF-calibration tone is also shown near 2.0~GHz (purple line).

In the RF spectrum four or more peaks are observed for each mode type, instead of a single peak expected from axisymmetric isotropic silicon simulations. We further investigate this behavior using three-dimensional numerical simulations that take into account both the in-plane anisotropy of silicon's~\cite{Hopcroft:2010hj} and the fluctuations in the $\wring$ throughout the cavity's perimeter. We found that the material anisotropy and a tiny variation of 10~nm (0.5\%) in $\wring$  is enough to account for the typical 10~MHz splitting between peaks within a given mode family (see Supplemental Material). 

The mechanical mode fitting and calibration give typical mechanical quality factors of $Q_\text{m}=2300$ at room temperature which are also confirmed by pump-probe experiments (see Supplemental Material), and optomechanical coupling rates as high as $g_0=23$~kHz for the type-I mode shown in Fig.~\ref{fig:experiments}e. The $g_0$ values are in good agreement with the simulated values when silicon's anisotropic elasticity is taken into account (see Supplementary Fig.~\ref{fig:aniso}b)). Fig.~\ref{fig:experiments}f shows the corresponding spectral density of displacement fluctuations ($S_{xx}$) for a the type-I mode with largest transduction, revealing a displacement sensitivity, given by the background noise, on the order of $5\times10^{-18}$m/$\sqrt{\text{Hz}}$ for a typical input optical power of $P_{in}=250~\mu$W.

In order to show the efficiency of the bullseye grating structure in preventing mechanical loss, we compare the measured mechanical spectra –obtained from a series of devices with varying values of $a$ – with the mechanical modes obtained  from the axisymmetric numerical model of the bullseye cavity. Fig.~\ref{fig:grating}a show the normalized (see \emph{Methods}) measurement results for both type-I and type-II modes as well as the band-edge frequency related to the third and fourth bands on Fig.~\ref{fig:crystal}d (dashed lines).  The corresponding linear crystal’s DOS is superimposed to the calculated mechanical modes on Fig.~\ref{fig:grating}b. Neither the axisymmetric or linear models capture the in-plane anisotropy of silicon nor any $\wring$ azimuthal fluctuations. Nonetheless those simulations reproduce most of the features observed on the measured spectra; inside the bandgap the calculated optomechanical coupling is as high as $30$~kHz and the mechanical mode is mostly confined within the $\wring$ (Fig.~\ref{fig:grating}c), resulting in high mechanical quality factors. Outside the bandgap, the once confined mechanical mode can couple to grating band modes and leak energy through the pedestal. This effect is clearly seen for $a=1.2~\mu$m in Fig. \ref{fig:grating}a; despite the large calculated optomechanical coupling of $27$~kHz, the mechanical mode spreads inside the circular grating resulting in a small signal noise ratio transduction signal. On the other hand, even inside the shaded region of Fig. \ref{fig:grating}b, the mechanical modes might not couple to the grating bands and result both in high optomechanical coupling and mechanical quality factors; the $a=1.0~\mu$m spectra is one such example. These results show a path to design a device with high mechanical quality factor and optomechanical coupling by adopting a simple one-dimensional model for the band structure and floating rings to infer the mechanical modes frequencies.

In summary, we present a new design for an optomechanical bullseye cavity with an independent confining approach for optical and mechanical modes. Such strategies allows for tailorable mechanical modes up to $8$~GHz and with $g_0$ as large as $200~$kHz. We experimentally demonstrate the bullseye optomechanical cavity using a standard optical lithography fabrication technique in a commercial CMOS foundry, resulting on modes between $2$~GHz and $3$~GHz with mechanical quality factors as high as $2300$ at room temperature. The whispering gallery optical modes presented quality factors as high as $40\times10^3$ and optomechanical coupling rate on the order of $g_0=23$~kHz, close to what is expected at these frequencies. Coupling between a single mechanical mode and multiple optical modes are also shown which could be used for efficient wavelength conversion~\cite{Hill:2012cka}. In addition the CMOS flexibility and the ability to independently tailor both optical and mechanical modes could be explored in the design of large arrays of coupled oscillators, enabling a route towards the investigation of topological optomechanical effects. 

\tocless{\section*{Methods}}

\tocless{\subsection*{Fabrication}}
The devices were fabricated through the EpiXfab initiative at IMEC on a silicon-on-insulator wafer (top silicon layer of 220~nm over 2~$\mu$m of buried silicon oxide). Simple disks were first patterned on the top silicon layer through deep UV lithography and plasma etching. A similar (aligned) patterning cycle was then performed to make the circular grooves, although now the plasma would only etch 150~nm of silicon instead of the whole 220~nm layer. Finally, an in-house post-process step was performed to selectively and isotropically remove the buried oxide using a diluted hydrofluoric acid in order to mechanically release the bullseye cavities.

\tocless{\subsection*{Experimental Protocol}}
In order to couple light to the optical modes confined at the bullseye’s outer ring, the taper-cavity distance was slowly decreased through a sub-micron positioning system while the optical transmission was simultaneously monitored in a broad wavelength range ($\approx100$~nm). The reduced overlap between the taper's guided mode and inner bullseye modes ensures that we are not accessing inner modes when the taper is far from the cavity, since all optical modes have alike optical quality factors. After such mode identification, the taper is positioned closer to the cavity to increase the optical coupling and hence the phase/frequency to amplitude transduction provided by the cavity.

\tocless{\subsection*{Mechanical spectra normalization}}
The detected current spectra is related to the mechanical noise spectra as~\cite{Gorodetsky:2010jd}:
\begin{equation}
 S_\text{II}(\Omega)= \frac{g_0^2K(\Omega)}{\xzpf^2\Omega^2}S_{xx}(\Omega) + \frac{\phi_0^2}{2}\frac{K(\Omega)}{\text{ENBW}}\delta(\Omega-\Omega_{mod})+ S_\text{shot}(\Omega),
\end{equation}
where $K(\Omega)$ is the cavity transduction function, $S_\text{shot}(\Omega)$ is the detector shot-noise, $\phi_0$ is the imprinted phase on the laser by the electro-optical phase modulator, $\text{ENBW}$ is the noise-equivalent band-width of the electrical spectrum analyzer and $\Omega_{mod}$ is the phase-modulation frequency. Figure \ref{fig:grating}a shows a series of mechanical spectra which were normalized by the phase modulator calibration tone (not shown) such that each curve is proportional to $g_0^2/\xzpf^2 S_{xx}(\Omega)$. After normalization each spectra is plotted in the same vertical dB scale.

\tocless{\section*{Author Contributions}}
F.G.S.S., Y.A.V.E., G.S.W. and T.P.M.A. designed the devices. F.G.S.S. performed the measurements with support from G.O.L., R.S.B. and supervision by T.P.M.A.. F.G.S.S., G.S.W. and T.P.M.A. analysed the measured data. All authors contributed to the writing of the manuscript.

\tocless{\section*{Acknowledgments}}
The authors would like to acknowledge Paulo Dainese for fruitful discussions and CCS-UNICAMP for providing the micro-fabrication infrastructure. This research was funded by the Sao Paulo State Research Foundation (FAPESP) (grants 2012/17610-3, 2012/17765-7 and 2013/06360-9), the National Counsel of Technological and Scientific Development (CNPQ - 550504/2012-5), and the Coordination for the Improvement of Higher Education Personnel (CAPES).

\onecolumngrid
\setcounter{figure}{0}
\setcounter{table}{0}
\setcounter{equation}{0}
\setcounter{section}{0}
\renewcommand{\theequation}{S\arabic{equation}}
\renewcommand{\thesection}{S\arabic{section}}
\renewcommand{\thesubsection}{\Alph{subsection}}
\renewcommand{\thesubsubsection}{\roman{subsubsection}}
\renewcommand{\thefigure}{S\arabic{figure}}
\renewcommand{\thetable}{S\arabic{table}}

\newpage

\tocless{\section*{Supplementary Information}}
\tableofcontents
\normalsize

\section{The linear crystal approximation}

\begin{figure}[htb!]
	\centerline{\includegraphics[scale=0.9]{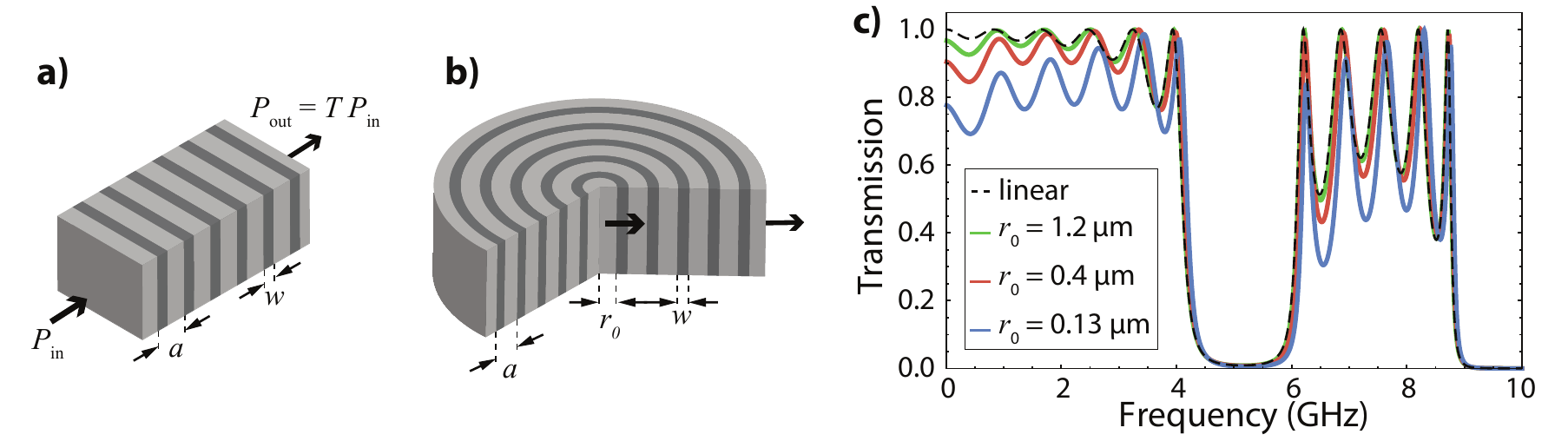}}
	\caption{\textbf{The linear grating approximation.}
		\textbf{a)} Linear and \textbf{b)} circular gratings used to illustrate the validity of the linear crystal approximation in predicting bandgaps. Material properties: $\rho_1=2329$~kg/m$^3$, $\rho_2=0.001\rho_1$, $c_{11}=165.6$~GPa, $c_{44}=79.5$~GPa, $v_{L,j}=\sqrt{c_{11}/\rho_j}$, $v_{S,j}=\sqrt{c_{44}/\rho_j}$ ($j=1,2$); geometric parameters: $a=1000$~nm, $w/a=0.2815$.
		\textbf{c)} Transmission spectra for linear and circular gratings (for varying $r_0$) calculated through the transfer-matrix method. Such spectra clearly show that the linear approximation works better in the short-wavelength limit, where $\omega r_0/v_\text{L}\gg1$.}
	\label{fig:linapprox}
\end{figure}

In this section we further justify the linear crystal approximation. The main idea behind it is that the curvature of cylindrically symmetric grating interfaces will not affect small wavelength radially-polarized waves and will be scattered like plane waves by plane interfaces. 
In order to better understand such approximation, we compare both plane and circular gratings infinite along the $z$-direction, like shown in Fig.~\ref{fig:linapprox}a-b). These structures may be connected to the proposed slab-like ones by an effective medium approximation: the bulk longitudinal, $v_L$, and shear, $v_S$, velocities of the infinite media are rescaled to reproduce the corresponding slab velocities of the etched grating regions. Such procedure is analogue to the effective index approximation often used in optical waveguides~\cite{Marcatili:2013fj, Okamoto:2006wc}. We shall focus on longitudinal waves since these are more closely approximate the bullseye's breathing modes.

Elastic normal modes in an isotropic solid may be described by the particle velocity field, $\vec{v}(\vec{r})$, and its eigenvalue equation~\cite{Auld:1990ub}
\begin{equation}
v_\text{L}^2\nabla(\nabla\cdot\vec{v}(\vec{r}))-v_\text{S}^2\nabla\times\left(\nabla\times\vec{v}(\vec{r})\right)+\omega^2\vec{v}(\vec{r})=0
\label{eq:part_veloc}
\end{equation}
where $v_\text{L}$ ($v_\text{S}$) is the bulk velocity of longitudinal (shear) waves and $\omega$ is the normal mode's frequency. Therefore, for longitudinal plane waves traveling along the $x$-direction, the solution to eq.~\eqref{eq:part_veloc} is
\begin{equation}
\vec{v}_\text{rect}=\left(A e^{i\omega x/v_\text{L}}+B e^{-i\omega x/v_\text{L}}\right)\hat{x}
\label{eq:rect_veloc}
\end{equation}
whereas cylindrically symmetric longitudinal waves in the $r$-direction may be written as
\begin{equation}
\vec{v}_\text{circ}=\left(AH_1^{(1)}\left(\frac{\omega r}{v_\text{L}}\right)+BH_1^{(2)}\left(\frac{\omega r}{v_\text{L}}\right)\right)\hat{r}
\label{eq:cyl_veloc}
\text{,}
\end{equation}
where $H_1^{(1)}$ and $H_1^{(2)}$ are first-order Hankel functions of the first and second kind, respectively, which asymptotically become cylindrical waves for $\omega r/v_\text{L}\gg1$.

Both $\vec{v}_\text{rect}$ and $\vec{v}_\text{circ}$ are of the form $A\vin+B\vout$, the coefficients $A$ and $B$ depending on boundary conditions while $\vin$ and $\vout$ describe counter-propagating waves. Furthermore, the solutions have constant value (by construction) along interfaces matching their symmetry. Therefore, a transfer-matrix approach may be used to calculate the transmission of elastic waves through a finite number of homogeneous layers such as in Fig.~\ref{fig:linapprox}a-b)~\cite{Yeh:1978ju}. Such spectrum is suitable for locating the structure's bandgaps because waves within it will be strongly reflected even by a finite grating. Our goal is then reduced to evaluating the range of parameters for which the linear and circular gratings have similar transmission spectra.

The transfer-matrix approach relies on writing the velocity field in every medium (labeled $j$) as $v_j=A_j\vin+B_j\vout$ and matching the boundary conditions (continuity of particle velocity and normal force~\cite{Auld:1990ub}) on each interface. For example, such procedure allows us to express $A_2$ and $B_2$ as linear combinations of $A_1$ and $B_1$,

\begin{equation*}
\begin{pmatrix}
A_2\\ B_2
\end{pmatrix}
=\tm_{21}
\begin{pmatrix}
A_1\\ B_1
\end{pmatrix}
\text{,}
\end{equation*}

where the square matrix $\tm_{21}$ is the so called transfer-matrix, which will depend on the symmetry of the problem.

Extending this argument to $N$ media,
\begin{equation}
\begin{pmatrix}
A_N\\ B_N
\end{pmatrix}
=\tm_{N,N-1}\ldots\tm_{32}\tm_{21}
\begin{pmatrix}
A_1\\ B_1
\end{pmatrix}
\text{.}
\label{eq:tmm}
\end{equation}

Assuming lossless media and denoting the incident and reflected waves by $\vin$ and $\vout$, respectively, energy conservation further requires $|A_1|^2=|B_1|^2+|A_N|^2$; $B_N=0$ is expected since there can be no reflected wave traveling within the last medium. The transmission spectrum is then calculated by evaluating the ratio $|A_N/A_1|^2$ for every frequency of interest.

Fig.~\ref{fig:linapprox}c) shows the transmission spectra calculated by this transfer-matrix approach for the linear and circular grating. As the radius of the smallest interface, $r_0$, decreases, it is clearly seen that the linear approximation starts to fail. On the other hand the transmission through the linear and circular gratings always disagree for low frequencies even for higher $r_0$ values. These results summarize our initial thesis: the linear grating suitably approximates the circular one as long as the wavelength is small compared to the layers' radius.

From eq.~\eqref{eq:rect_veloc} and \eqref{eq:cyl_veloc}, such conclusions can be made more quantitative. The validity of the linear approximation depends mainly on how well cylindrical waves approximate the Hankel functions that analytically solve the problem. Therefore, the linear crystal approximation is expected to work better for higher values of $\omega r/v_\text{L}$.

\tocless{\subsection*{Calculation of transfer-matrices}}
Here we show the details about the calculations of the transfer-matrices for linearly and cylindrically symmetric problems. Such calculations rely on the boundary conditions of continuity of the velocity field, $\vec{v}$, and of the normal force given by $\ten{T}\cdot\hat{n}$, $\ten{T}$ representing the stress tensor.

The stress tensor may be expressed as a function of the generalized Hooke's law as $\ten{T}=\ten{C}:\ten{S}$, where $\ten{C}$ is the 4-th rank stiffness tensor and $\ten{S}$ is the strain tensor, also given by the symmetric gradient of the displacement field, $\ten{S}=\nabla_S\vec{u}$. For normal modes, the harmonic temporal behavior implies that $\vec{v}=-i\omega\vec{u}$.

For the planar interface described by $v_\text{rect}$ (eq.~\eqref{eq:rect_veloc}), the boundary conditions become continuity of $v_\text{rect}$ and of $\ten{T}\cdot\hat{x}$ at a interface plane, $x=w$, between media 1 and 2. We therefore get the transfer-matrix
\begin{equation}
	\tm_{21}=
	\begin{pmatrix}
	\frac{ v_\text{L,1} \rho_1+v_\text{L,2} \rho_2}{2 v_\text{L,2} \rho_2}
	\exp\left\lbrace\frac{2 i w \omega }{v_\text{L,1}}-\frac{i (v_\text{L,1}+v_\text{L,2}) w \omega }{v_\text{L,1} v_\text{L,2}}\right\rbrace &
	\frac{v_\text{L,2} \rho_2-v_\text{L,1} \rho_1}{2 v_\text{L,2} \rho_2}
	\exp\left\lbrace -\frac{i (v_\text{L,1}+v_\text{L,2}) w \omega }{v_\text{L,1} v_\text{L,2}}\right\rbrace\\
	\frac{v_\text{L,2} \rho_2-v_\text{L,1} \rho_1}{2 v_\text{L,2} \rho_2}
	\exp\left\lbrace\frac{2 i w \omega }{v_\text{L,1}}+\frac{i (v_\text{L,1}-v_\text{L,2}) w \omega }{v_\text{L,1} v_\text{L,2}}\right\rbrace &
	\frac{v_\text{L,1} \rho_1+v_\text{L,2} \rho_2}{2 v_\text{L,2} \rho_2}
	\exp\left\lbrace\frac{i (v_\text{L,1}-v_\text{L,2}) w \omega }{v_\text{L,1} v_\text{L,2}}\right\rbrace \\
	\end{pmatrix}
	\label{eq:plane_tm}
\end{equation}
where $\rho_j$ and $v_\text{L,j}$ are respectively the density and bulk longitudinal velocity in medium $j$.

Now for the cylindrically symmetric problem, we use the continuity of $v_\text{circ}$ from eq.~\eqref{eq:cyl_veloc} and the corresponding normal force, $\ten{T}\cdot\hat{r}$ at a cylindrical surface $r=a$ separating media 1 and 2, to obtain

\begin{equation*}
\tm_{21}=\frac{i\pi}{4 v_\text{L,2}^2 \rho_2}
\begin{pmatrix}
d_1 & t_{12} \\
t_{21} & d_2
\end{pmatrix}
\end{equation*}

with matrix elements given by

\begin{footnotesize}
\begin{align*}
d_1 &= H_1^{(1)}\left(\frac{a \omega }{v_\text{L,1}}\right) \left\lbrace a v_\text{L,2} \rho_2 \omega H_0^{(2)}\left(\frac{a \omega }{v_\text{L,2}}\right)+2 \left(v_\text{S,1}^2 \rho_1-v_\text{S,2}^2 \rho_2\right) H_1^{(2)}\left(\frac{a \omega }{v_\text{L,2}}\right)\right\rbrace-a v_\text{L,1} \rho_1 \omega H_0^{(1)}\left(\frac{a \omega }{v_\text{L,1}}\right) H_1^{(2)}\left(\frac{a \omega }{v_\text{L,2}}\right) \\
d_2 &= H_1^{(1)}\left(\frac{a \omega }{v_\text{L,2}}\right) \left\lbrace a v_\text{L,1} \rho_1 \omega H_0^{(2)}\left(\frac{a \omega }{v_\text{L,1}}\right)+2 \left(v_\text{S,2}^2 \rho_2-v_\text{S,1}^2 \rho_1\right) H_1^{(2)}\left(\frac{a \omega }{v_\text{L,1}}\right)\right\rbrace -a v_\text{L,2} \rho_2 \omega H_0^{(1)}\left(\frac{a \omega }{v_\text{L,2}}\right) H_1^{(2)}\left(\frac{a \omega }{v_\text{L,1}}\right) \\
t_{12} &= a v_\text{L,2} \rho_2 \omega  H_0^{(2)}\left(\frac{a \omega}{v_\text{L,2}}\right) H_1^{(2)}\left(\frac{a \omega }{v_\text{L,1}}\right)+\left\lbrace 2 \left(v_\text{S,1}^2 \rho_1-v_\text{S,2}^2 \rho_2\right) H_1^{(2)}\left(\frac{a \omega }{v_\text{L,1}}\right)-a v_\text{L,1} \rho_1 \omega  H_0^{(2)}\left(\frac{a \omega }{v_\text{L,1}}\right)\right\rbrace H_1^{(2)}\left(\frac{a \omega}{v_\text{L,2}}\right) \\
t_{21} &= -a v_\text{L,2} \rho_2 \omega  H_0^{(1)}\left(\frac{a \omega }{v_\text{L,2}}\right) H_1^{(1)}\left(\frac{a \omega }{v_\text{L,1}}\right) - \left\lbrace 2 \left(v_\text{S,1}^2 \rho_1-v_\text{S,2}^2 \rho_2\right) H_1^{(1)}\left(\frac{a \omega}{v_\text{L,1}}\right)-a v_\text{L,1} \rho_1 \omega  H_0^{(1)}\left(\frac{a \omega }{v_\text{L,1}}\right)\right\rbrace H_1^{(1)}\left(\frac{a \omega }{v_\text{L,2}}\right) \\
\end{align*}
\end{footnotesize}
where now the shear velocities $v_\text{S,j}$ explicitly appear in the transfer-matrix in contrast to eq.~\eqref{eq:plane_tm}. This can be understood from the evaluation of the strain field, since the gradient of the radial displacement field $\vec{u}(r)$ also has a $\varphi$-component for these cylindrically symmetric solutions.

\section{Pump-probe mechanical spectroscopy}

\begin{figure*}[htb!]
	\centerline{\includegraphics[scale=0.9]{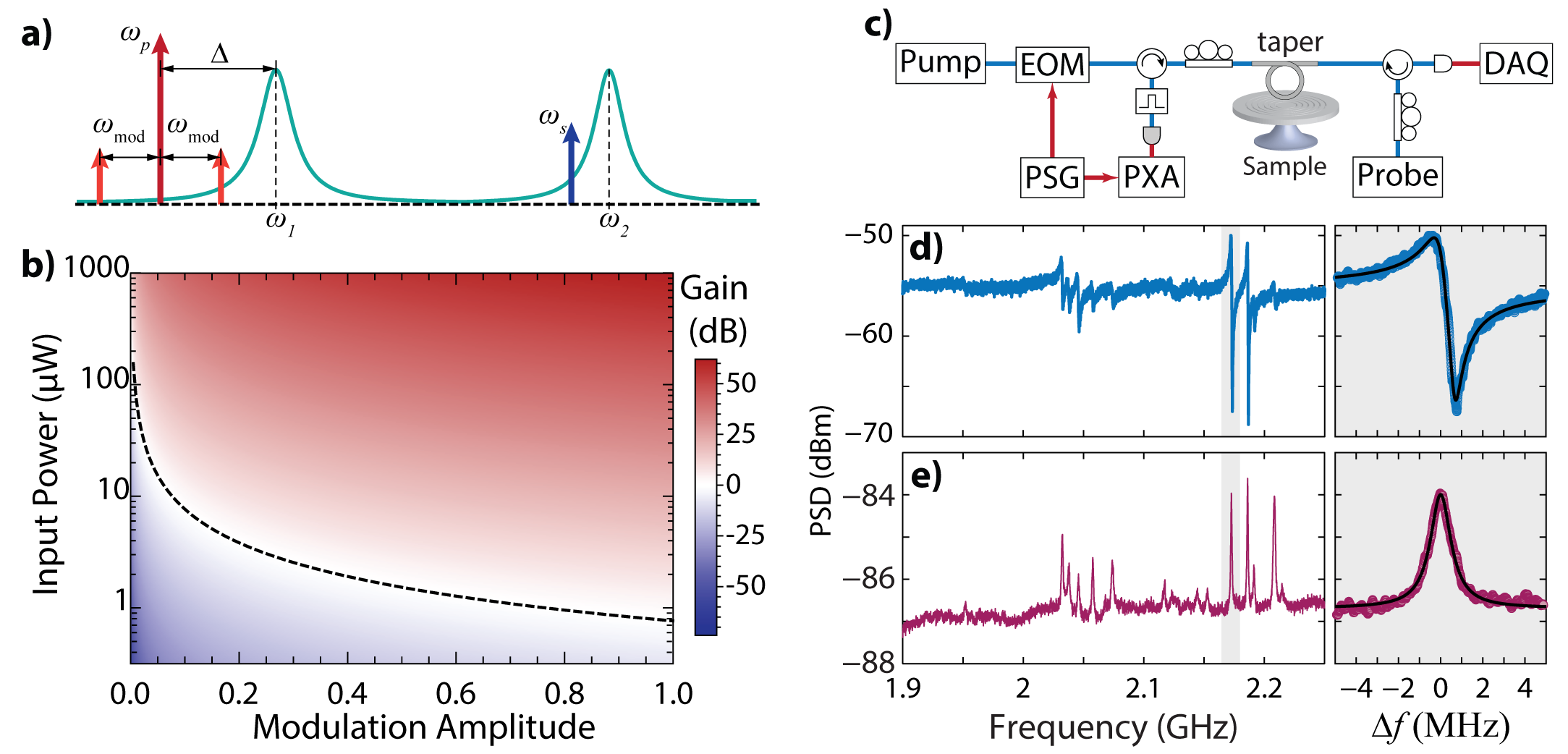}}
	\caption{\textbf{Pump-probe measurements.}
		\textbf{a)} An amplitude modulated pump laser (red) locked to the half-linewidth of an optical resonance harmonically drives the mechanical mode through radiation pressure force; a probe laser (blue) tuned to an optical resonance with equal transverse profile but a few free spectral ranges away, reads the driven motion through the optomechanical interaction.
		\textbf{b)} The signal gain compared to the thermal driven motion strongly depends on the modulation amplitude and input power of the pump laser. The dashed line shows the unity gain, above which the pump-probe scheme is actually advantageous.
		\textbf{c)} In order to avoid contamination of the probe signal by the pump, the beams are launched in opposite directions and isolated from each other through optical circulators. A bandpass filter tuned to the probe band further improves the degree of isolation.
		\textbf{d)} A Fano lineshape arises from interference between the driven mechanical motion and the broadband Kerr effect. Such lineshape may be fitted to yield the mechanical mode's frequency and quality factor.
		\textbf{e)} The pump-probe measurement agrees to the thermally driven one; although having a similar signal-to-noise ratio, the pump-probe spectrum has a much high extinction.}
	\label{fig:pump-probe}
\end{figure*}

A way to characterize an harmonic oscillator is by looking at its response due to a harmonic driving force. In optomechanical systems, this can be done to probe the mechanical modes with improved signal-to-noise ratio~\cite{VanLaer:2015jk}. To do so, we used a pump-probe scheme as illustrated in Fig.~\ref{fig:pump-probe}a). A strong amplitude modulated pump laser drives the mechanical mode through radiation pressure. The mechanical oscillation turns into modulation of the optical resonance frequency, which can be measured in the transmission of a weak probe laser tuned to an optical mode with similar transverse profile but a few free spectral ranges away from the pumped mode; this strategy is necessary to isolate the probe signal from amplitude modulation coming directly from the pump laser.

The gain in signal relative to the thermally driven mechanical spectrum can be quantified by comparing the displacement noise spectrum in both cases. The thermal spectrum is given by the well known expression~\cite{Aspelmeyer:2014ce}

\begin{equation}
S_{xx}^\text{(thermal)}(\Omega)=\frac{2\gamma\kB T/\meff}{(\omegam^2-\Omega^2)^2+(\gamma\Omega)^2}
\label{eq:thermal_noise}
\end{equation}

Now to find the displacement noise spectrum for the optically driven case, we solve the equations of motion for the pumped optical mode's amplitude, $a$, and for the excited mechanical mode's displacement, $x$,

\begin{align}
\dot{a} &=-(i\Delta+\frac{\kappa}{2})a-i G x a+\sqrt{\kappae}\ain(1+\delta\cos\omegamod t)
\label{eq:opt_dyn}\\
\ddot{x}&+\gamma\dot{x}+\omegam^2 x=\frac{\hbar G}{\meff}|a|^2
\label{eq:mech_dyn}
\text{,}
\end{align}
where $G=g_0/\xzpf$. The last term in eq.~\eqref{eq:opt_dyn} accounts for the modulated pump laser, where $|\ain|^2$ is the incident photon flux (the input pump power reads $\hbar\omegap|\ain|^2$ where $\omegap$ is the pump laser frequency), $\omegamod$ the modulation frequency, $\delta$ describes the modulation depth and $\kappae$ is the extrinsic loss rate. The coupling terms describe the optical frequency shift in eq.~\eqref{eq:opt_dyn} and radiation pressure in \eqref{eq:mech_dyn}. The remaining parameters concern the modes regardless of optomechanical interaction: $\Delta$ is the pump laser's detuning to the optical resonance, $\kappa$ is the total optical loss rate, $\omegam$ is the mechanical resonance frequency, $\gamma$ the mechanical loss rate, $\meff$ the mechanical mode's effective mass and $\hbar$ is the reduced Planck constant.

The dominant terms of the solution to eq. \eqref{eq:opt_dyn} and \eqref{eq:mech_dyn} should be either static or oscillating with the modulation frequency. Therefore,
\begin{align*}
a&=\alpha_0+\alpha_{-}e^{i\omegamod t}+\alpha_{+}e^{-i\omegamod t}\\
x&=x_0+\frac{1}{2}(x_1e^{i\omegamod t}+x_1^*e^{-i\omegamod t})
\end{align*}

By keeping the pump power low so that dynamical backaction may be ignored and by absorbing static optical resonance shifts in $\Delta$, one gets
\begin{equation}
x_1=\frac{2\hbar G}{\meff} \frac{\alpha_0^*\alpha_{-}+\alpha_0\alpha_{+}^*}{\omegam^2-\omegamod^2+i\gamma\omegamod}
\end{equation}
with
\begin{align}
\alpha_0&=\frac{\kappae\ain}{i\Delta+\kappa/2}\\
\alpha_-&=\frac{\kappae\ain\delta}{i(\Delta+\omegamod)+\kappa/2}\\
\alpha_+&=\frac{\kappae\ain\delta}{i(\Delta-\omegamod)+\kappa/2}
\end{align}

The displacement noise spectrum associated to this solution would show an infinitely narrow peak at $\omegamod$, which when integrated should lead to the driven variance
\begin{equation}
\Delta x^2=\mean{(x-\mean{x})^2} =\frac{|x_1|^2}{2}
\text{.}
\end{equation}

Apart from transduction constants, the actually measured peak has its width limited both by the modulator's linewidth and by the spectrum analyzer's Effective Noise BandWidth (ENBW). In this experiment we used an electro-optical modulator driven by a sub-Hz linewidth radio-frequency signal generator, whereas an ENBW of $\approx100$~kHz was enough to resolve the mechanical resonances. Therefore, the peak's width was always limited by ENBW and such peak intensity as a function of $\omegamod$ reads

\begin{align}
S_{xx}^\text{(driven)}(\omegamod)&=\frac{2\pi}{\text{ENBW}}\Delta x^2=\nonumber\\
\frac{4\pi}{\text{ENBW}} \left(\frac{\hbar G}{\meff}\right)^2 & \frac{\left|(\alpha_0^*\alpha_{-}+\alpha_0\alpha_{+}^*)\right|^2}{(\omegam^2-\omegamod^2)^2+(\gamma\omegamod)^2}
\label{eq:driven_noise}
\end{align}

It is worth noting that, although similar experimental setups are used for probing the thermal and driven noises, the driven noise acquisition requires using the spectrum analyzer actually as a demodulator tuned to the modulator's frequency $\omegamod$.

Now the gain in signal obtained by this pump-probe strategy may be easily expressed from eq.~\eqref{eq:thermal_noise} and \eqref{eq:driven_noise}:

\begin{equation}
\text{gain}=\frac{2\pi}{\text{ENBW}}\frac{\left|\hbar G(\alpha_0^*\alpha_{-}+\alpha_0\alpha_{+}^*)\right|^2}{\gamma\meff\kB T}
\label{eq:gain}
\end{equation}

It immediately follows from eq.~\eqref{eq:gain} that the gain should benefit both from a high input pump power and from a large modulation depth. It also follows that the pump-probe measurement is most effective when the pump laser is tuned to the half-linewidth of the pumped resonance (see Fig.~\ref{fig:pump-probe}a)) and that it is limited to the cavity's bandwidth $\kappa$.

Fig.~\ref{fig:pump-probe}b) shows a plot of the gain as a function of the input power and modulation depth for cavity parameters matching our devices (see Fig.~\ref{fig:experiments} and main text). The black dashed line shows the curve of unity gain, above which the pump-probe method is indeed advantageous.

We performed pump-probe measurements following the setup of Fig.~\ref{fig:pump-probe}c). An electro-optical modulator driven by a RF-signal generator modulates the pump laser which is then guided towards the cavity. In order to avoid the thermo-optical instability, the pump laser is actively locked to the blue side of an optical resonance at half-linewidth. The counter-propagating probe laser is launched from the opposite direction and tuned to the half-linewidth of an optical mode of the same transverse family but a few free-spectral-range (FSR) apart ($\approx20$~nm). Optical circulators and a bandpass filter ensure that the photocurrent seen by the electrical spectrum analyzer is only due to the probe transmission. Under typical experimental conditions, pump power around $100~\mu$W and modulation depth of $5$~\%, the gain would be approximately $25$~dB according to Fig.~\ref{fig:pump-probe}.

The measured pump-probe spectrum is a bit more complicated than eq.~\eqref{eq:driven_noise} due to other optical nonlinearities. Broadband nonlinear effects such as Kerr, free carrier dispersion and free carrier absorption interfere with the narrow-band optomechanical effect into a Fano lineshape (Fig.~\ref{fig:pump-probe}d)). These extra nonlinearities may be lumped into a frequency independent complex parameter $k_\text{NL}$, so that we may fit the Fano resonances to the expression
\begin{equation}
S_{II}(\omegamod)=\left|k_\text{NL}+\frac{k_\text{OM}}{\omegam^2-\omegamod^2+i\gamma\omegamod}\right|^2
\end{equation}
which allows for calculating the quality factor of each mechanical mode; the Q-factors calculated by such approach agree to those calculated from the thermal noise spectrum shown in Fig.~\ref{fig:experiments} of the main text.

The main advantage of the pump-probe method in these devices is the much larger extinction near mechanical resonances compared to the thermal spectrum, as shown in Fig.~\ref{fig:pump-probe}d-e). Such large extinction allows for easily locating the mechanical modes which can be further studied (including $g_0$ measurements) in the direct detection setup of Fig.~\ref{fig:experiments}b).

\section{Effects of eccentricity and anisotropy on the bullseye's mechanical modes}

In this section we show how silicon's elastic anisotropy and possible fluctuations in $\wring$ lead to a more thorough comprehension of the mechanical spectra shown in Fig.~\ref{fig:experiments}.

\begin{figure*}[htb!]
	\centerline{\includegraphics[scale=0.9]{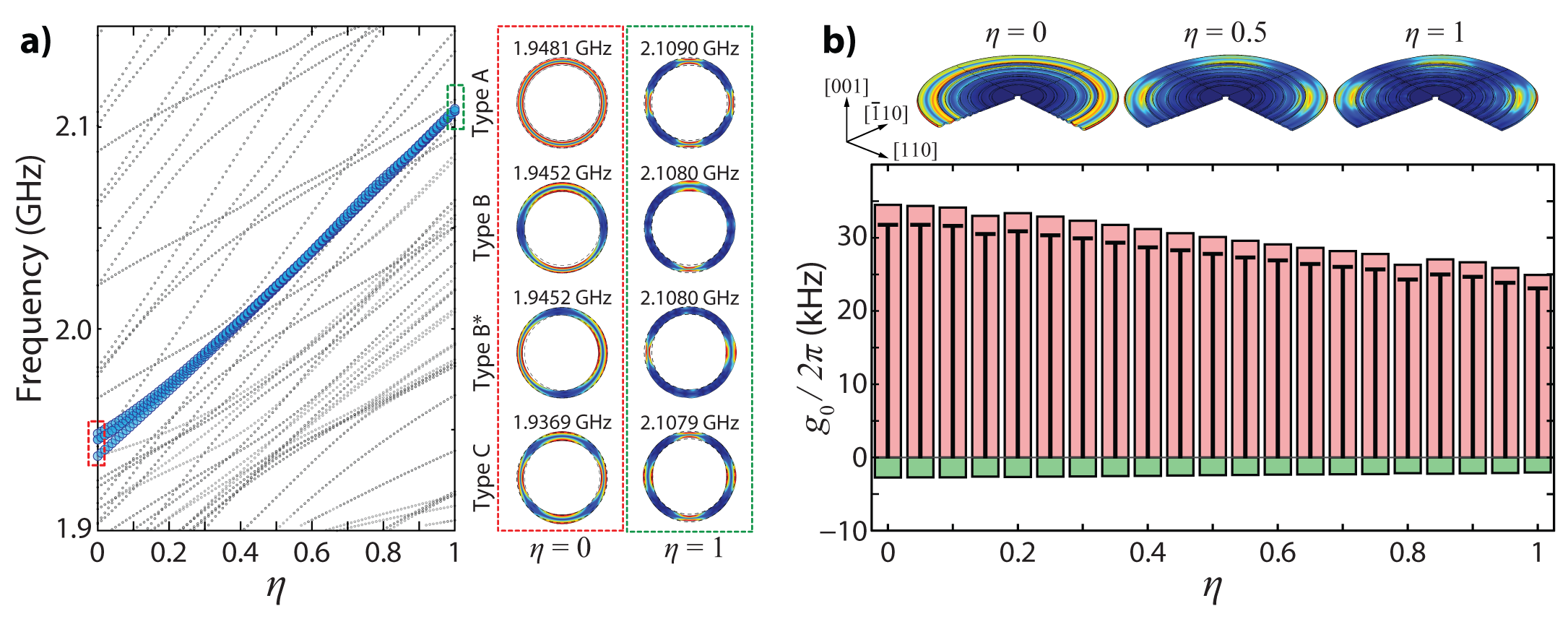}}
	\caption{\textbf{Effect of anisotropy on mechanical modes.}
		\textbf{a)} A three-dimensional FEM simulation shows the breathing mode and 3 extra mechanical modes for an isotropic approximation ($\eta=0$), all of which couple to the ring-like optical mode at similar optomechanical coupling rates. These 4 modes are hybridized as silicon's anisotropy is taken into account ($\eta>0$, such that the exact anisotropy is approached as $\eta$ approaches 1). The measured splitting of nearly $10$~MHz, however, may be better understood as a combination between anisotropy and fluctuations in $\wring$ (see Fig.~\ref{fig:eccentricity}).
		\textbf{b)} Numerical calculations for the breathing mode shows that the total optomechanical coupling rate $g_0$ (black) diminishes towards the measured values as anisotropy is taken into account following the trend of the dominant photoelastic contribution (red); the smaller moving boundary contribution (green) barely changes.}
	\label{fig:aniso}
\end{figure*}

To do so, we first perform three dimensional finite element simulations of the ring structure (Fig.~\ref{fig:aniso}a)) focusing on quasi-radially polarized modes. In order to clearly understand how the anisotropy affects the modes, we write silicon's stiffness tensor:

\begin{equation*}
C(\eta)=
\begin{pmatrix}
	c_{11} & c_{12} & c_{12} & 0      & 0      & 0      \\
	c_{12} & c_{11} & c_{12} & 0      & 0      & 0      \\
	c_{12} & c_{12} & c_{11} & 0      & 0      & 0      \\
	0      & 0      & 0      & c_{44}^*(\eta) & 0      & 0      \\
	0      & 0      & 0      & 0      & c_{44}^*(\eta) & 0      \\
	0      & 0      & 0      & 0      & 0      & c_{44}^*(\eta) \\
\end{pmatrix}
\end{equation*}

where

\begin{equation*}
c_{44}^*(\eta)=\eta c_{44}+(1-\eta)\frac{c_{11}-c_{12}}{2}
\end{equation*}
$\eta$ being a simulation parameter that continuously switches from an isotropic approximation ($\eta=0$) to the real anisotropic stiffness tensor for silicon ($\eta=1$); $c_{11}=165.6$~GPa, $c_{12}=63.9$~GPa and $c_{44}=79.5$~GPa are the stiffness constants for anisotropic silicon.

\begin{figure*}[hbt!]
	\centerline{\includegraphics[scale=0.9]{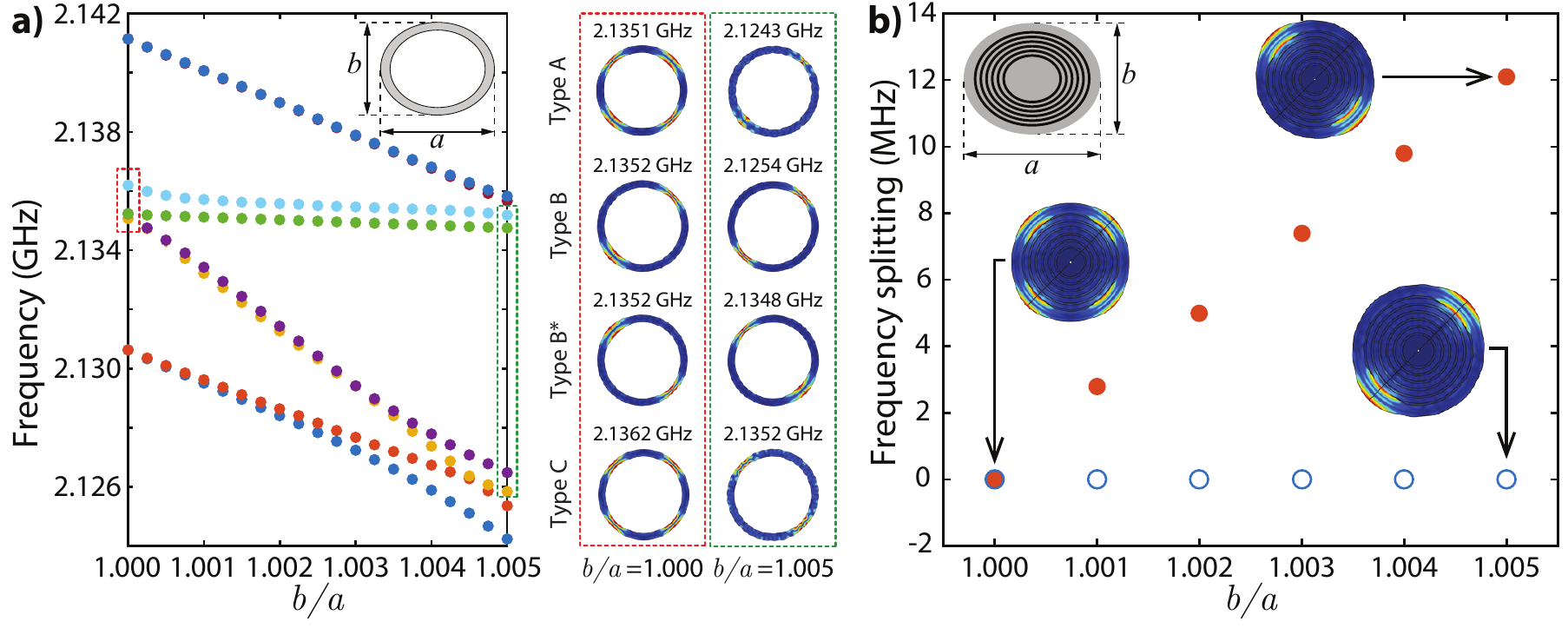}}
	\caption{\textbf{Effect of eccentricity on mechanical modes.}
		\textbf{a)} Simulated mechanical dispersion as a function of $b/a$ for a silicon ring (\emph{inset}) with anisotropic stiffness. A splitting of $\sim10$~MHz arises as $b$ exceeds $a$ by 0.5\%. The eccentricity-induced mode hybridization is such that the former type A mode gives origin to two branches with similar mode profiles, hence similar optomechanical couplings.
		\textbf{b)} A similar splitting behavior also shows up for bullseye simulations (performed for half-structure with symmetric boundary conditions). Because the real fluctuations are random, such mechanism may be responsible for the multiple peaks observed within the RF spectra.}
	\label{fig:eccentricity}
\end{figure*}

The three-dimensional isotropic simulation ($\eta=0$) shows 4 modes with large optomechanical coupling to the outer ring optical mode: the azimuthally symmetric breathing mode (type A) and three extra ones that break the azimuthal symmetry, two of which (type B and B$^*$) are degenerate and connected by $\pi/2$ rotations. As $\eta$ is raised to 1, the 4 modes get more hybridized and their splitting actually becomes smaller; also, the twofold degeneracy is not broken since the anisotropic stiffness (or more generally, silicon's cubic structure) is still symmetric under $\pi$ rotations (Fig.~\ref{fig:aniso}a)). On the other hand, the optomechanical coupling rate for the Type A mode gets closer to the measured value $g_0/2\pi=23.1\pm0.2$~kHz as $\eta$ approaches 1 due to the further localization of mechanical strain around the ring region, which suppresses the photoelastic effect (Fig.~\ref{fig:aniso}b)). However, the B, B$^*$ and C modes still have very small $g_0$.

There remains the problem of understanding the multiple peaks within each family. A quick estimation would show that a splitting of 10~MHz in $\omegam\approx2$~GHz could be caused by a fluctuation of 10~nm on $\wring=2~\mu$m; controlling such geometric fluctuations is far beyond the limits of deep UV lithography.

To further investigate whether such $\wring$ fluctuations are a reasonable explanation for the observed 10~MHz splitting, one has to show not only that the splitting arises from such fluctuations, but that these also cause more modes to have a high $g_0$. Therefore, we performed another set of three-dimensional finite element simulations of ring structures, now assuming the more realistic anisotropic stiffness tensor $\ten{C}(\eta=1)$ and varying the eccentricity of the devices as shown in Fig.~\ref{fig:eccentricity}~ by keeping $a=\wring$ and varying $b$; this is the most simple way of simulating such fluctuations without recurring to more complicated stochastic methods while still getting a lot of physical insight.

\begin{figure*}[ht!]
	\centerline{\includegraphics[scale=0.9]{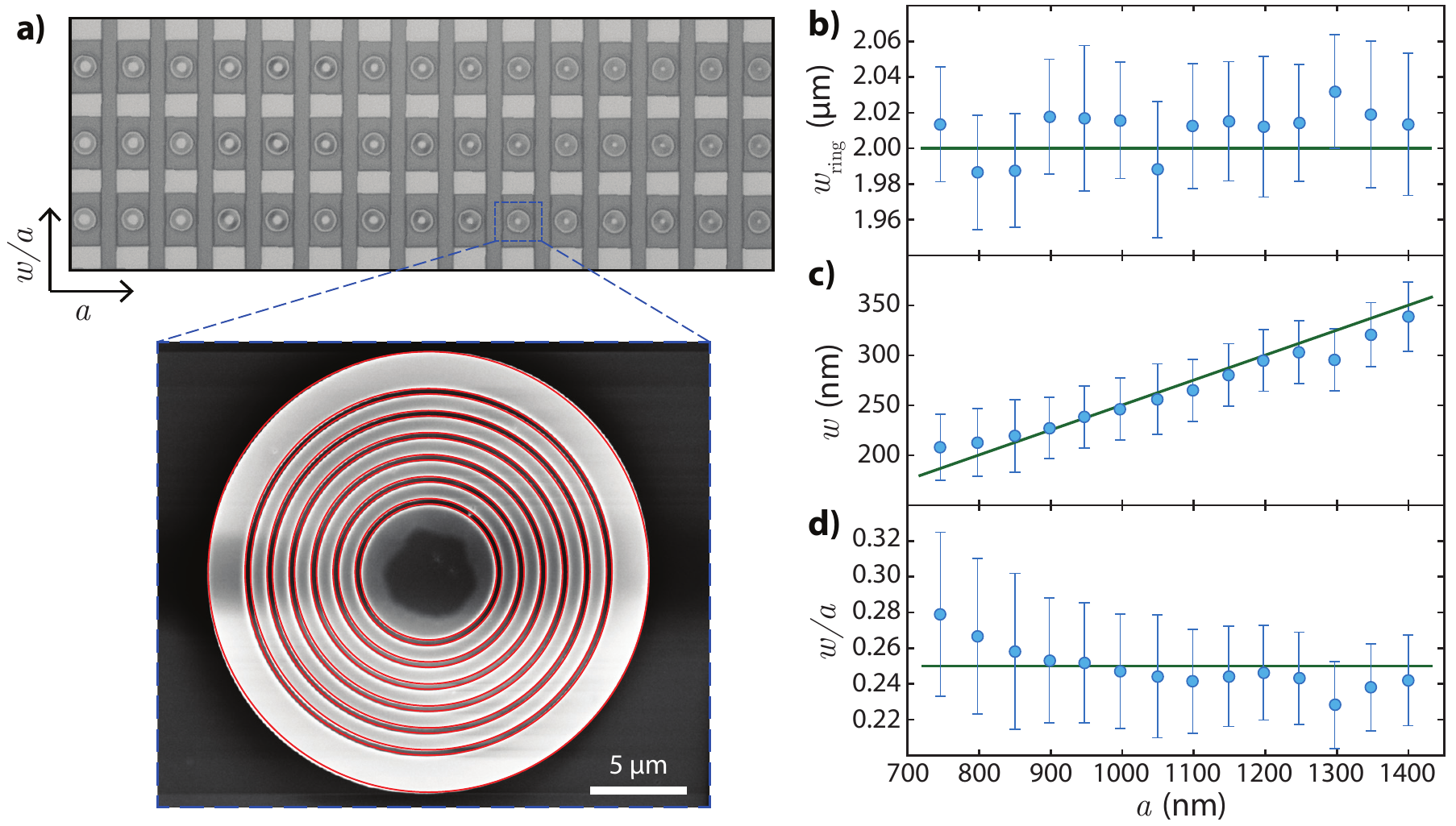}}
	\caption{\textbf{Scaling the bullseye disk fabrication.}
		\textbf{a)} An array of bullseye cavities is shown with constant $\wring=2.0~\mu$m and varying $a$ and $w$. The central row is the optomechanically characterized one (see Fig.~\ref{fig:experiments}). SEM images of the lower row devices were fitted (red curves) as shown in the inset in order to check the geometrical parameters.
		\textbf{b)} The ring-width $\wring$ matches the nominal value within the fitting error, as well as the
		\textbf{c)} groove width $w$. The fitting uncertainty is dominated by systematic error, $\pm2$ pixels, intrinsic to our fitting algorithm.
		\textbf{d)} $w/a$ also matches the nominal values within error bars. The higher error bars for smaller $w/a$ values is due to the relative importance of the $\pm$ 2 pixels of systematic error compared to both $w$ and $a$.}
	\label{fig:geom}
\end{figure*}

\begin{figure*}[ht!]
	\centerline{\includegraphics[scale=0.9]{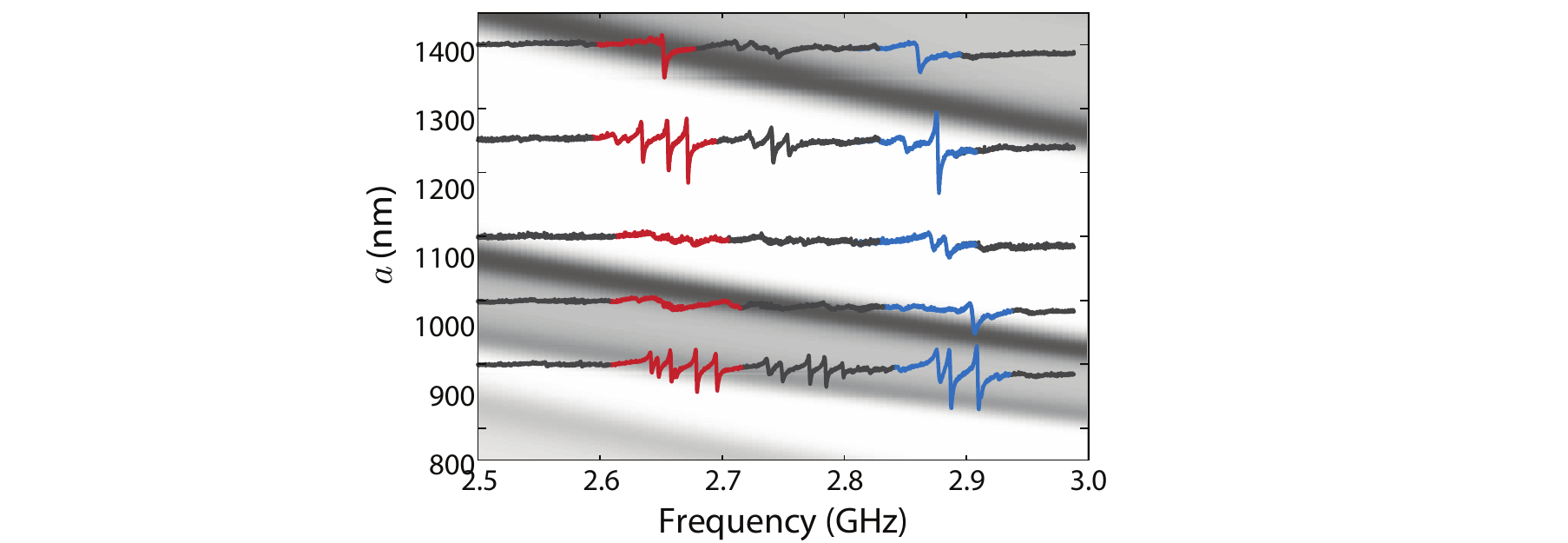}}
	\caption{\textbf{Mechanical modes for $\boldsymbol{\wring=1.5~\mu}$m.}
		Pump-probe spectra as function of $a$ for bullseye disks with $\wring=1.5~\mu$m and $w/a=0.28$; the mechanical mode frequencies along with extinction and quality factor approximately matches the linear crystal DOS (gray scale), like observed for $\wring=2.0~mu$m.}
	\label{fig:fano2}
\end{figure*}

Fig.~\ref{fig:eccentricity}a) shows the ring dispersion as a function of $b/a$. It is clearly seen that the mode hybridization caused by eccentricity leads to splittings on the order of 10~MHz. Additionally, from the highlighted mode profiles, it would be expected two modes with large $g_0$ (those where expansions or compressions are always in phase) instead of only one for the perfectly circular device.

Such conclusions are reproduced when investigating the actual bullseye disk as shown in Fig.~\ref{fig:eccentricity}b), whose simulations were performed for half-bullseye with symmetric boundary conditions along a diameter. The Type A mode basically breaks up in two branches whose separation reaches 12~MHz as $b/a$ approaches 1.005. Again, both branches should show similar optomechanical couplings. Finally, such considerations can be extended to explain a higher number of peaks within each family by recalling that the $\wring$ variations intrinsic to the fabrication process are of a random nature.

\section{Scalable and tailorable of bullseye optomechanical cavities}

In order to check whether the actual dimensions match the nominal values, we fitted SEM images of the bullseye devices. The images were acquired in a NOVA 200 Nanolab Dual Beam (FIB-SEM) System and later processed to correct residual astigmatism. We rescale each SEM image to ensure that the diameter measured along the image’s x- and y-directions are equal. The validity of such scaling is confirmed by comparing the eccentricity impact on the mechanical mode splitting (Fig. S4), suggesting that diameter fluctuations should be lower than 1\% of the $\wring$ width (20 nm). Also, we calibrate the absolute distance scale by assuming that each disk diameter is equal to the nominal one $24~\mu$m. 

All size measurements are within the nominal value to the fitting error --- our fitting algorithm has an uncertainty of $\pm15$~nm ($\pm2$ pixels), which is much larger than statistical errors for every fitted image. These results demonstrate the scalability of bullseye optomechanical disks to a precision of 15~nm in CMOS-compatible Foundries, an important step towards commercial applications of optomechanical resonators.

Finally, we demonstrate mechanical tailorability by exploring the dependence of the mechanical frequency on $\wring$.  To do so, we characterized a set of bullseye devices with $\wring=1.5~\mu$m which are expected to have mechanical resonances related to Type I and Type II modes slightly below 3~GHz according to Fig.~\ref{fig:ring}. Fig.~\ref{fig:fano2} shows pump-probe measurements for such set of devices over the linear crystal DOS. Like for those devices with $\wring=2.0~\mu$m, the linear crystal DOS and mechanical resonances agree (except for a small displacement in $a$) and again two main families of resonances are noticed, corresponding to Type I and Type II mechanical mode profiles.

\footnotesize
\bibliographystyle{naturemag}
\vskip -0.075in
\tocless{{\small\bibliography{bullseye.bib}}}

\end{document}